\documentclass[aps,twocolumn]{revtex4}

\usepackage{graphics}

\begin{document}

\bibliographystyle{apsrev}

\title{The LISA Response Function}
\author{Neil J. Cornish and Louis J. Rubbo}
\affiliation{Department of Physics, Montana State University, Bozeman, MT 59717}

\begin{abstract}
The orbital motion of the Laser Interferometer Space Antenna (LISA) introduces
modulations into the observed gravitational wave signal. These modulations can be used to
determine the location and orientation of a gravitational wave source. The complete LISA
response to an arbitrary gravitational wave is derived using a coordinate free approach in
the transverse-traceless gauge. The general response function reduces to that found by
Cutler\cite{cc} for low frequency, monochromatic plane waves. Estimates of the noise in
the detector are found to be complicated by the time variation of the interferometer arm
lengths.
\end{abstract}


\maketitle

\section{Introduction}
In this paper we derive the response of the Laser Interferometer Space Antenna (LISA) to
an arbitrary gravitational wave.  Our detector response function includes the full
orbital motion and is valid at all frequencies.  Previous treatments have either assumed
that LISA is stationary with respect to the background sky\cite{aet,lhh,cl}, or have been
limited to the low frequency limit\cite{cc,mh}.

The LISA mission\cite{lppar} calls for three spacecraft to orbit the Sun in an
equilateral triangular formation. The center-of-mass for the constellation, known as the
guiding center, follows a circular orbit at 1 AU and has an orbital period of one year.
In addition to the bulk motion of the detector about the Sun, the triangular formation
will cartwheel about the guiding center in a clockwise manner as seen by an
observer at the Sun. The orbital motion introduces frequency (Doppler), amplitude, and
phase modulation into the observed gravitational wave signal. These effects have been
calculated in the low frequency limit, where the antenna pattern is well approximated by
a quadrupole, and the Doppler modulation is due to the guiding center motion. At higher
frequencies the antenna pattern becomes more complicated, and the ``rolling'' Doppler
modulation due to the cartwheel motion has to be included.

The divide between high and low frequencies roughly coincides with the divide between
gravitational waves with wavelengths shorter or longer than the arms of the detector.
To be precise, the dividing line is the transfer frequency $f_\ast \equiv c/(2\pi L)$,
where $L$ is the unperturbed distance between spacecraft\cite{lhh}.
For LISA, with its mean arm length of $5 \times 10^{6}$ km, the transfer frequency has an
approximate value of $10$ mHz. Above the transfer frequency the antenna pattern is
distinctly non-quadrapolar\cite{njc}. The frequencies at which the guiding center and rolling
motion impart measurable effects are easily estimated. Since both the guiding center and
cartwheel motions have periods of one year, the Doppler modulations will enter as sidebands
separated by the modulation frequency $f_m = 1/{\rm year}$. Equating the modulation frequency to the
Doppler shift, $\delta f \simeq (v/c)f$, for motion with velocity $v$  yields the
characteristic frequency $f_v=c f_m/v$ at which Doppler modulation becomes measurable.
Assuming $5 \times 10^{6}$ km armlengths the cartwheel turns with
velocity $v/c = 0.192\times 10^{-5}$, while the guiding center moves
with velocity $v/c = 0.994\times 10^{-4}$. Thus,
the Doppler modulation due to the guiding center's motion becomes measurable
at frequencies above $f_{gc} = 0.3$ mHz, while the rolling cartwheel motion becomes
important above $f_{r} = 16 $ mHz. This demonstrates that at low
frequencies only the bulk motion of the detector needs to be
considered, while at high frequencies the cartwheel motion also
needs to be included.

Our calculations are performed using the coordinate free approach introduced in
Ref.\cite{cl} (see also Ref.~\cite{ron} for a closely related approach).
This allows our results to
be applied to any variation on the current
LISA design, or to any follow on mission. The low frequency limit of
our general detector response function yields a simple result that can be shown to
agree with Cutler's\cite{cc}.
Throughout this paper we use natural units where $G=c=1$, however we will report all
frequencies in terms of Hertz.

\section{Detector response}

The detector response to a gravitational wave source located in
the $\hat{n}$ direction can be found using Barycentric coordinates
$(t,{\bf x})$ and the transverse-traceless gauge to describe a
plane gravitational wave ${\bf h}(q,\widehat\Omega)$ propagating
in the $\widehat\Omega=-\hat{n}$ direction. The surfaces of
constant phase are given by $\xi=t+\hat{n}\cdot {\bf x}={\rm
const.}$. A general gravitational wave can be decomposed
into two polarization states:
\begin{equation}\label{genh}
{\bf h}(\xi,\hat{n}) = h_{+}(\xi){\bf e}^{+} + h_{\times}(\xi){\bf e}^{\times}
\end{equation}
where ${\bf e}^{+}$ and ${\bf e}^{\times}$ are the polarization tensors
\begin{eqnarray}\label{eten}
{\bf e}^{+} &=& \hat{u}\otimes \hat{u} - \hat{v}\otimes \hat{v},  \nonumber \\
{\bf e}^{\times} &=& \hat{u}\otimes \hat{v} + \hat{v}\otimes \hat{u}.
\end{eqnarray}
The basis vectors $\hat{u}$, $\hat{v}$ and the source location $\hat{n}$
can be expressed in terms of the location of the source $(\theta,\phi)$ according to
\begin{eqnarray}\label{wave}
&&\hat{u} = \cos\theta\cos\phi \, \hat{x} +\cos\theta\sin\phi \,
\hat{y} -\sin\theta \, \hat{z} \nonumber \\
&&\hat{v} = \sin\phi \, \hat{x} -\cos\phi \, \hat{y} \nonumber \\
&&\hat{n} = \sin\theta \cos\phi \, \hat{x} + \sin \theta\sin\phi \, \hat{y}+\cos\theta
\, \hat{z}\, .
\end{eqnarray}

Following the Doppler tracking calculations described in Ref.\cite{cl}, we find that
the optical path length between spacecraft $i$ and spacecraft $j$ may be written as
\begin{eqnarray}\label{distance}
\ell_{ij}(t_i) &=& \int_i^j \sqrt{ g_{\mu\nu} dx^\mu dx^\nu} \nonumber \\
&=& \vert {\bf x}_j(t_j) - {\bf x}_i(t_i) \vert \nonumber \\
&&\; + \frac{1}{2} (\hat{r}_{ij}(t_i) \otimes
\hat{r}_{ij}(t_i)) : \int^{j}_{i} {\bf h}(\xi(\lambda)) \, d\lambda \, .
\end{eqnarray}
Here $g_{\mu\nu}=\eta_{\mu\nu}+h_{\mu\nu}$ and $\vert {\bf x} -{\bf y} \vert$
denotes the Cartesian distance between ${\bf x}$ and ${\bf y}$.
The unit vector
\begin{equation}
\hat{r}_{ij}(t_i) = \frac{{\bf x}_j(t_j) - {\bf x}_i(t_i)}{\ell_{ij}(t_i)} \, 
\end{equation}
points from spacecraft $i$ at the time of
emission, $t_i$, to spacecraft $j$ at the time of reception, $t_j$.
Finally, the quantity $\xi(\lambda)$ is the parameterized wave
variable
\begin{equation}
\xi(\lambda) = t(\lambda) - \hat{\Omega} \cdot {\bf x}(\lambda) \, .
\end{equation}
Explicitly, the time and position depend on the parameterization
in the following way
\begin{eqnarray}
t(\lambda) &=& t_i +\lambda \\
{\bf x}(\lambda) &=& {\bf x}_{i}(t_i) + \lambda \,
\hat{r}_{ij}(t_i) \, .
\end{eqnarray}
The variation in the Cartesian distance between the spacecraft can be separated into
a contribution due to the Sun and other Solar system bodies, and a small perturbation
due to the gravitational wave. Denoting the unperturbed spacecraft locations by ${\bf x}^0(t)$
and integrating the geodesic equation for the metric $g_{\mu\nu}$ yields
\begin{eqnarray}
&& \vert {\bf x}_j(t_j) - {\bf x}_i(t_i) \vert = \vert {\bf x}^0_j(t_j) - {\bf x}^0_i(t_i) \vert
\nonumber \\
&& \quad + \hat{r}_{ij}(t_i)\cdot \int^j {\bf h}(t-\hat{\Omega}\cdot{\bf x}_j(t))\cdot 
\frac{d{\bf x}_j(t)}{d t} \, dt \nonumber \\
&& \quad - \hat{r}_{ij}(t_i)\cdot \int^i {\bf h}(t-\hat{\Omega}\cdot{\bf x}_i(t))\cdot 
\frac{d{\bf x}_i(t)}{d t} \, dt \, .
\end{eqnarray}
The gravitational wave dependent terms in the above equation can be ignored as they
are down by a factor of
$v = \vert d{\bf x}(t)/d t \vert \simeq 10^{-4}$ compared to the leading order
gravitational wave contribution described in (\ref{distance}). Thus we may write
\begin{equation}
\ell_{ij}(t_i) = \ell^0_{ij}(t) + \delta \ell_{ij}(t_i)
\end{equation}
where
\begin{equation}
\ell^0_{ij}(t_i) = \vert {\bf x}^0_j(t_j) - {\bf x}^0_i(t_i) \vert  \, ,
\end{equation}
and
\begin{equation}
\delta \ell_{ij}(t_i) = \frac{1}{2} \frac{\hat{r}_{ij}(t_i)
\otimes \hat{r}_{ij}(t_i)}{1 - \hat{\Omega}\cdot\hat{r}_{ij}(t_i)}
: \int^{\xi_j}_{\xi_i} {\bf h}(\xi) \, d\xi \, .
\end{equation}
Here we have used
\begin{equation}
\frac{{\rm d}\xi(\lambda)}{{\rm d}\lambda} = 1 -
\hat{\Omega}\cdot\hat{r}_{ij}(t_i)
\end{equation}
to make a change of variables in the integration.
Defining {\bf H}($a$,$b$) to be the antiderivative of the
gravitational wave
\begin{equation}
{\bf H}(a,b) \equiv \int^a_b {\bf h}(\xi) \, d\xi
\end{equation}
simplifies our expression further to
\begin{equation}\label{main}
\delta \ell_{ij}(t_i) = \frac{1}{2} \frac{\hat{r}_{ij}(t_i)
\otimes \hat{r}_{ij}(t_i)}{1 + \hat{n}\cdot\hat{r}_{ij}(t_i)} :
{\bf H}(\xi_j,\xi_i) \, .
\end{equation}
where we have used the relationship $\hat{\Omega} = -\hat{n}$. To leading order
in ${\bf h}$, the time of reception $t_j$ is defined in terms of the time of
emission $t_i$ by the implicit relation
\begin{equation}
\ell^0_{ij}(t_i)
          = \vert {\bf x}^0_j(t_i+\ell^0_{ij}(t_i)) - {\bf x}^0_i(t_i) \vert \, .
\end{equation}

The gravitational wave can be decomposed into frequency components
\begin{equation}
{\bf h}(\xi) = \int_{-\infty}^{\infty} \tilde{{\bf h}}(\omega) e^{i \omega \xi} d\omega \, ,
\end{equation}
which allows us to write
\begin{equation}
{\bf H}(a,b) = \int_{-\infty}^{\infty} \frac{\tilde{{\bf h}}(\omega)}{i \omega}
\left( e^{i\omega a} - e^{i\omega b} \right) d \omega
\end{equation}
and
\begin{equation}\label{mainf}
\delta \ell_{ij}(t_i) = \ell_{ij}(t_i) \int_{-\infty}^{\infty}
{\bf D}(\omega,t_i,\hat{n}): \tilde{{\bf h}}(\omega)e^{i \omega \xi_i} d\omega
\end{equation}
where the one-arm detector tensor is given by
\begin{equation}
{\bf D}(\omega,t_i,\hat{n}) = \frac{1}{2} \left(\hat{r}_{ij}(t_i)
\otimes \hat{r}_{ij}(t_i)\right) {\cal T}(\omega,t_i,\hat{n})
\end{equation}
and the transfer function takes the form
\begin{eqnarray}
{\cal T}(\omega,t_i,\hat{n})&=& {\rm sinc}\left[\frac{\omega}{2\omega_{ij}}
(1+\hat{n} \cdot \hat{r}_{ij}(t_i))\right] \nonumber \\
& \times & \exp\left[i \frac{\omega}{2\omega_{ij}}
(1+\hat{n} \cdot \hat{r}_{ij}(t_i))\right]  \, .
\end{eqnarray}
Here $\omega_{ij}=1/\ell_{ij}(t_i)$ is the angular transfer frequency for the arm.

The connection between the optical path length variations and the detector output
depends on the interferometer design. The original proposal was to use laser transponders
at the end-stations to send back a phased locked signal. A more recent proposal is to
eliminate the transponders and turn LISA into a virtual interferometer where the signal
is put together in software. The raw ingredients for this procedure are the phase
differences between the received and transmitted laser light along each arm. The signal
transmitted from spacecraft $i$ that is received at spacecraft $j$ at time $t_j$ has its
phase compared to the local reference to give the output $\Phi_{ij}(t_j)$. The phase difference
has contributions from the laser phase noise, $C(t)$, optical path length variations, shot
noise $n^s(t)$ and acceleration noise ${\bf n}^a(t)$:
\begin{eqnarray}\label{phs}
\Phi_{ij}(t_j) &=& C_i(t_i) - C_j(t_j) + 2\pi \nu_0\,(\delta \ell_{ij}(t_i)+\Delta \ell_{ij}(t_i))
\nonumber \\
&&+ n_{ij}^s(t_j)
-\hat{r}_{ij}(t_i)\cdot\left({\bf n}_{ij}^a(t_j)-{\bf n}_{ji}^a(t_i) \right).
\end{eqnarray}
Here $t_i$ is given implicitly by $t_i=t_j-\ell_{ij}(t_i)$ and $\nu_0$ is the laser frequency.
We have included the variations in the optical path length caused by
gravitational waves, $\delta \ell_{ij}(t_i)$, and those caused by orbital
effects, $\Delta \ell_{ij}(t_i)$. In what follows we will ignore the orbital contributions to
the phase shift as they can be removed by high pass filtering.
The subscripts on the noise sources
identify the particular component that is responsible: $C_i$ is the phase noise introduced by
the laser on spacecraft $i$, $n_{ij}^s$ denotes the shot noise in the photodetector on
spacecraft $j$ used to measure the phase of the signal from spacecraft $i$,
and ${\bf n}^a_{ij}$ denotes the noise introduced by the accelerometers on
spacecraft $j$ that are mounted on the
optical assembly that points toward spacecraft $i$.

The three LISA spacecraft will report six phase difference measurements which can then
be used to construct a variety of interferometer outputs. The simplest are the
three Michelson signals that can be formed by choosing one of the spacecraft as the
vertex and using the other two as end-stations. The Michelson signal extracted from
vertex 1 has the form
\begin{equation}\label{mic}
S_1(t) = \Phi_{12}(t_2)+\Phi_{21}(t)-\Phi_{13}(t_3)-\Phi_{31}(t) \, ,
\end{equation}
where $t_2$ and $t_3$ are given implicitly by
\begin{eqnarray}
t_2 &=& t - \ell_{21}(t_2) \nonumber \\
t_3 &=& t - \ell_{31}(t_3) \, .
\end{eqnarray}
Unfortunately, the Michelson signals will be swamped by laser
phase noise, so a more complicated virtual interferometer signal
has to be used. The $X$ variables are a set of three
Michelson-like signals that cancel the laser phase noise
\cite{ta}. The $X$ signal extracted from vertex 1 has the form
\begin{eqnarray}\label{x1}
X_1(t) &=& \Phi_{12}(t_2)-\Phi_{12}(t_1+t_2-t)  \nonumber \\
&+& \Phi_{21}(t)- \Phi_{21}(t_1)  \nonumber \\
&-& \Phi_{13}(t_3) + \Phi_{13}(t_1+t_3-t) \nonumber \\
&-& \Phi_{31}(t) + \Phi_{31}(t_1)  \, .
\end{eqnarray}
The times $t_1,\; t_2$ and $t_3$ are defined implicitly:
\begin{eqnarray}
t_1 &=& t - \ell_{12}(t_1)-\ell_{21}(t_2) \nonumber \\
    &=& t - \ell_{13}(t_1)-\ell_{31}(t_3) \nonumber \\
t_2 &=& t - \ell_{21}(t_2) \nonumber \\
t_3 &=& t - \ell_{31}(t_3) \, .
\end{eqnarray}
Given a gravitational wave signal ${\bf h}(q, \widehat\Omega)$, a model of the
instrument noise, and a description of the interferometer's orbit, we can
use equations (\ref{main}), (\ref{phs}) and (\ref{x1}) to calculate the detector response.
Since the entire calculation is performed in Barycentric coordinates, the time $t$ that
appears in (\ref{mic}) and (\ref{x1}) is not the time $\tau$ measured by the
clock on spacecraft 1. They are related by the standard time dilation formula $d\tau =
dt \sqrt{1-v_1^2(t)}$. However, since we only need to work to leading order in $v$, there
is no need to distinguish between $t$ and $\tau$.

Our expression for the LISA response is much more complicated than the previous
approximate descriptions. The time-variation of the optical path lengths is the main
cause of the difficulty. It is responsible for the implicit relations that riddle the
calculation. The path length variations have three main causes - intrinsic, tidal, and
pointing. The intrinsic variations are part and parcel of the cartwheel orbit, which only
keeps the distance between the spacecraft constant to leading order in the orbital
eccentricity. The tidal variations are caused by the gravitational pull of other solar
system bodies, most notably the Earth and Jupiter. The pointing corrections are a
relativistic effect caused by the finite propagation speed of the lasers, which means
that the spacecraft move between transmission and reception. The latter effect can be
separated from the others:
\begin{equation}\label{el12}
\ell_{ij}(t_i)=L_{ij}(t_i)\left[1+\hat{r}_{ij}(t_i)\cdot{\bf v}_{j}(t_i)+{\cal O}(v^2)\right]
\end{equation}
where ${\bf v}_{j}$ is the velocity of spacecraft $j$ and
\begin{equation}
L_{ij}(t_i) = \vert {\bf x}_{j}(t_i) - {\bf x}_{i}(t_i) \vert .
\end{equation}
Ignoring tidal distortions and working to second order in the orbital eccentricity $e$,
the orbits described in the Appendix yield
\begin{equation}\label{l12}
L_{12}(t) = L \left(1 + \frac{e}{32}\Bigl[ 15\sin(\alpha+\frac{\pi}{6})-\cos(3\alpha)
\Bigr]\right),
\end{equation}
and similar, yet slightly different, expressions for $L_{13}(t)$ and $L_{23}(t)$.
Here $\alpha(t)=2\pi f_m t +\kappa$ is the orbital phase and $f_m=1/{\rm year}$ is the
orbital frequency. The mean armlength $L$ is related to the eccentricity
$e$ and semi-major axis $a$ by
$L=2\sqrt{3} a e$. Setting $L= 5 \times 10^9$ meters yields $e=0.00965\approx 10^{-2}$.
The spacecraft have velocities of
order $v\approx 2\pi f_m a \approx 10^{-4}$. Using these numbers we see that the
lowest order intrinsic variation is far larger than the pointing variation. The tidal
variations turn out to be comparable to the intrinsic variation\cite{lppar} and
therefore should not be ignored.

\section{Static Limit}

As a point of reference, we can apply our general method to a static, equal arm
detector interacting with a monochromatic, plane-fronted
gravitational wave propagating in the $\widehat\Omega = -\hat n$
direction with principle polarization axes ${\bf p}$ and ${\bf q}$:
\begin{equation}\label{mono}
{\bf h}(f,\xi) = A_{+}e^{2\pi if(t+\hat{n}\cdot{\bf x})}{\mbox{\boldmath$\epsilon$}}^{+}
+ A_{\times} e^{2\pi if(t+\hat{n}\cdot{\bf x})}{\mbox{\boldmath$\epsilon$}}^{\times},
\end{equation}
where $A_{+}$ and $A_{\times}$ are complex constants and
\begin{eqnarray}
{\mbox{\boldmath$\epsilon$}}^{+} &=& \hat{p}\otimes\hat{p} - \hat{q}\otimes\hat{q} , \nonumber \\
{\mbox{\boldmath$\epsilon$}}^{\times} &=& \hat{p}\otimes\hat{q} + \hat{q}\otimes\hat{p} \, .
\end{eqnarray}
Defining the polarization angle $\psi=-{\rm arctan}(\hat{v}\cdot{\bf p}/\hat{u}\cdot{\bf p})$ we have
\begin{eqnarray}\label{polten}
{\mbox{\boldmath$\epsilon$}}^{+} &=&\cos 2\psi\, {\bf e}^{+} - \sin 2\psi\, {\bf e}^{\times}, \nonumber \\
{\mbox{\boldmath$\epsilon$}}^{\times} &=& \sin 2\psi\, {\bf e}^{+}
+ \cos 2\psi\, {\bf e}^{\times}.
\end{eqnarray}
Thus, in terms of the general decomposition (\ref{genh}) we have
\begin{eqnarray}
h_+(t) &=& \left(A_{+}\cos 2\psi + A_{\times}\sin 2 \psi\right)e^{2\pi if(t+\hat{n}\cdot{\bf x})}, 
\nonumber \\
h_\times(t) &=& \left(A_{\times}\cos 2\psi - A_{+}\sin 2 \psi\right)e^{2\pi if(t+\hat{n}\cdot{\bf x})}.
\end{eqnarray}

The signal portion of the $X$ variable defined in (\ref{x1}) reduces to
\begin{eqnarray}
&& X^s_1(t)=  {2\pi \nu_0} \Bigl( \delta \ell_{12}(t-2L) - \delta \ell_{12}(t-4L)  \nonumber \\
&& \quad + \delta \ell_{21}(t-L) - \delta \ell_{21}(t-3L) - \delta \ell_{13}(t-2L) \nonumber \\
&& \quad + \delta \ell_{13}(t-4L) - \delta \ell_{31}(t-L) + \delta \ell_{31}(t-3L) \Bigr) .
\end{eqnarray}
In terms of the strain $x_1(t)=X^s_1(t)/(2\pi L \nu_0)$ we have
\begin{equation}\label{xstrain}
x_1(t)={\bf D}(\hat{n},f):{\bf h}(f, \xi)\sin^2(f/f_*)
\end{equation}
where
\begin{eqnarray}
{\bf D}(\hat{n},f) &=& \frac{1}{2}\left( (\hat{r}_{12}\otimes\hat{r}_{12})
        \, {\cal T}(\hat{r}_{12}\cdot\hat{n},f) \right. \nonumber \\
       && - \left.
    (\hat{r}_{13}\otimes\hat{r}_{13})\, {\cal T}(\hat{r}_{13}\cdot\hat{n},f) \right)
\end{eqnarray}
and
\begin{eqnarray}
    && {\cal T}(s,f) = \frac{1}{2}\left[{\rm sinc}\left( \frac{f(1+s)}{2f_* }\right)
    \exp\left(-i\frac{f}{2f_*}(3 - s)\right) \right. \nonumber \\
    && \quad \quad + \left. {\rm sinc}\left(\frac{f(1-s)}{2f_* }\right) \exp\left(-i\frac{f}{2f_*}(1-
    s)\right)\right].
\end{eqnarray}
Orienting the detector in the $x-y$ plane according to Figure 2 of Ref.~\cite{cc}, we have
\begin{eqnarray}
&&\hat{r}_{12} = \cos (\pi / 12) \, \hat{x} + \sin (\pi / 12) \,
\hat{y} \nonumber \\
&&\hat{r}_{13} = \cos (5\pi / 12) \, \hat{x} + \sin (5\pi / 12) \, \hat{y} \, .
\end{eqnarray}
Combing these expression with equations (\ref{eten}) and (\ref{wave}) yields
\begin{eqnarray}\label{ds}
(\hat{r}_{12}\otimes\hat{r}_{12}):{\bf e}^{+} &=& \frac{1}{2} \left(
(1+\cos^{2}\theta)\sin(2\phi+\pi/3) -
\sin^{2}\theta \right) \nonumber \\
(\hat{r}_{12}\otimes\hat{r}_{12}):{\bf e}^{\times} &=&
\cos\theta \, \sin(2\phi-\pi/6)  \nonumber \\
(\hat{r}_{13}\otimes\hat{r}_{13}):{\bf e}^{+} &=& \frac{1}{2} \left(
(1+\cos^{2}\theta)\sin(2\phi-\pi/3) -
\sin^{2}\theta \right)  \nonumber \\
(\hat{r}_{13}\otimes\hat{r}_{13}):{\bf e}^{\times} &=& - \cos\theta \, \sin(2\phi+\pi/6).
\end{eqnarray}
The above collection of equations, (\ref{xstrain}) through (\ref{ds}), fully define the
detector response in the static limit. The $X$ variable response $x_1(t)$ is related to the
Michelson response $s_1(t)$ by
\begin{equation}\label{mstrain}
s_1(t) = x_1(t) \sin^{-2}(f/f_*) = {\bf D}(\hat{n},f):{\bf h}(f, \xi)\, .
\end{equation}
Our expression (\ref{mstrain}) for $s_1(t)$ agrees with that quoted in Ref.\cite{cl}.
In the low frequency limit, $f \ll f_*$, the transfer function reduces to unity, ${\cal T}=1$, and
\begin{equation}
s_1(t) = \Bigl(A_+ F^{+}(\theta,\phi,\psi) + A_\times F^{\times}(\theta,\phi,\psi)\Bigr)e^{2\pi i f t}.
\end{equation}
The beam pattern factors
\begin{eqnarray}
F^{+}(\theta,\phi,\psi) &=& \frac{1}{2} (\hat{r}_{12} \otimes \hat{r}_{12} - \hat{r}_{13}
\otimes \hat{r}_{13}):{\mbox{\boldmath$\epsilon$}}^{+} \nonumber\\
F^{\times}(\theta,\phi,\psi) &=& \frac{1}{2} (\hat{r}_{12} \otimes \hat{r}_{12} -
\hat{r}_{13} \otimes
\hat{r}_{13}):{\mbox{\boldmath$\epsilon$}}^{\times}
\end{eqnarray}
take their familiar form:\cite{kip}
\begin{eqnarray}
&&F^{+} = \frac{\sqrt{3}}{2} \Bigl( \frac{1}{2} (1+\cos^{2}\theta) \cos 2\phi  \cos 2\psi \nonumber \\
&& \hspace*{0.8in} - \cos\theta  \sin 2\phi  \sin 2\psi \Bigr) \nonumber \\
&&F^{\times}  = \frac{\sqrt{3}}{2} \Bigl( \frac{1}{2} (1+\cos^{2}\theta) \cos
2\phi  \sin 2\psi \nonumber \\
&& \hspace*{0.8in} + \cos\theta  \sin 2\phi \cos 2\psi \Bigr).
\end{eqnarray}
Notice that the overall factor of $\sqrt{3}/2$ compared to a detector with $90^{o}$ arms
come out naturally in our calculation.

\section{Low frequency limit}

As a further point of reference, we can apply our general result to the low frequency
limit considered in Ref.\cite{cc}. The low frequency limit is defined by the condition
$f\ll f_*$, where $f_*\equiv1/(2\pi L)$ is the typical transfer frequency along each arm.
A LISA mission with $L= 5 \times 10^9$ meter arms has a transfer frequency of $f_* =
0.00954 \approx 10^{-2}$ Hz. The motion of the LISA constellation is included to leading
order in the eccentricity, and the gravitational wave is taken to be monochromatic,
plane-fronted and propagating in the $\widehat\Omega = -\hat n$ direction:
\begin{equation}
{\bf h}(t,{\bf x})=A_+ {\mbox{\boldmath$\epsilon$}}^+ \cos(2\pi f(t+\hat{n}\cdot{\bf x}))+ A_\times
{\mbox{\boldmath$\epsilon$}}^\times \sin(2\pi f(t+\hat{n}\cdot{\bf x})) \, .
\end{equation}
The two orthogonal polarizations have constant real amplitudes $A_+$ and $A_\times$. The basic
Michelson signal considered by Cutler\cite{cc} takes the form
\begin{eqnarray}
s_1(t) &=& \frac{\delta\ell_{12}(t-2L)+\delta\ell_{21}(t-L)}{2L} \nonumber \\
&-& \frac{\delta\ell_{13}(t-2L)
+\delta\ell_{31}(t-L)}{2L} \, .
\end{eqnarray}
This expression ignores the time variation of the armlengths due to higher order terms in
the orbital eccentricity or perturbations from other solar system bodies.
Using (\ref{main}) we find
\begin{eqnarray}
s_1(t) &=& F^+(t) A_+ \cos(2\pi f [t+ \hat{n}\cdot {\bf x}_1(t)]) \nonumber \\
&+& F^\times(t) A_\times \sin(2\pi f [t+ \hat{n}\cdot {\bf x}_1(t)]) ,
\end{eqnarray}
where
\begin{eqnarray}
F^+(t) &=& \frac{1}{2}\left(\cos 2\psi \, D^+(t) - \sin 2\psi \, D^\times(t) \right)\nonumber\\
F^\times(t) &=& \frac{1}{2}\left(\sin 2\psi \, D^+(t) + \cos 2\psi \, D^\times(t) \right),
\end{eqnarray}
and
\begin{eqnarray}
D^+(t) &\equiv& \left(\hat{r}_{12}(t)\otimes \hat{r}_{12}(t)-\hat{r}_{13}(t)\otimes
\hat{r}_{13}(t)\right):{\bf e}^+ \nonumber \\
D^\times(t) &\equiv& \left(\hat{r}_{12}(t)\otimes \hat{r}_{12}(t)-\hat{r}_{13}(t)\otimes
\hat{r}_{13}(t)\right):{\bf e}^\times .
\end{eqnarray}
The expression for the strain in the detector can be re-arranged using double
angle identities to read:
\begin{equation}\label{lf}
s_1(t) = A(t) \cos \left[ 2\pi f t + \phi_D(t) + \phi_P(t) \right]\, .
\end{equation}
The amplitude modulation $A(t)$, frequency modulation $\phi_D(t)$ and
phase modulation $\phi_P(t)$ are given by
\begin{eqnarray}
A(t) &=& \left[ (A_+ F^+(t))^2+(A_\times F^\times (t))^2 \right]^{1/2} \\
\nonumber \\
\phi_D(t) & = & 2\pi f \hat{n}\cdot {\bf x}_1(t) \nonumber \\
&=& 2\pi f a \sin\theta \cos(\alpha - \phi) \\
\nonumber \\
\phi_P(t) & = & - {\rm arctan}\left(\frac{A_\times F^\times(t)}{A_+ F^+(t)}\right) \, .
\end{eqnarray}
Using the orbits described in the Appendix, the coordinates of each spacecraft are
given to leading order in the eccentricity by
\begin{eqnarray}\label{xyz}
x &=& a\cos(\alpha) + ae\left(\sin\alpha\cos\alpha\sin\beta
-(1+\sin^2\alpha)\cos\beta\right)\nonumber \\
y &=& a\sin(\alpha) + ae\left(\sin\alpha\cos\alpha\cos\beta
-(1+\cos^2\alpha)\sin\beta\right)\nonumber \\
z & = & - \sqrt{3} a e \cos(\alpha-\beta) \, ,
\end{eqnarray}
where $\alpha = 2\pi f_m t + \kappa$ is the phase of the guiding center and
$\beta=2n\pi/3+\lambda$ is the relative phase of each spacecraft in the constellation
$(n=0,1,2)$. The unit vectors $\hat{r}_{ij}(t)$ can be derived from the coordinates given
in (\ref{xyz}). Putting this all together yields
\begin{eqnarray}
D^+(t) &=& \frac{\sqrt{3}}{64} \Big[ -36 \sin^2\theta \sin(2\alpha(t)-2\lambda )
\nonumber\\
&+& (3+\cos 2\theta)\Big(\cos 2\phi (9 \sin 2\lambda-\sin(4\alpha(t)-2\lambda))
\nonumber\\
&& \quad +\sin 2\phi(\cos (4\alpha(t)-2\lambda) - 9\cos 2\lambda )\Big)
\nonumber\\
&-& 4 \sqrt{3} \sin 2\theta \Big(\sin(3\alpha(t)-2\lambda-\phi)\nonumber\\
&& \quad - 3\sin(\alpha(t)-2\lambda+\phi) \Big) \Big]
\end{eqnarray}
and
\begin{eqnarray}
D^\times(t) &=& \frac{1}{16} \Big[ \sqrt{3} \cos \theta \Big(9\cos (2\lambda-2\phi)
\nonumber\\
&& \quad - \cos(4\alpha(t) - 2\lambda-2\phi)\Big) \nonumber\\
&-&  6\sin\theta\Big(\cos(3\alpha(t)-2\lambda-\phi) \nonumber\\
&& \quad + 3\cos(\alpha(t)-2\lambda+\phi)\Big) \Big]
\end{eqnarray}
Finally, for circular Newtonian binaries, the polarization angle $\psi$ can be related to the
angular momentum orientation $\hat{L}\rightarrow (\theta_L,\phi_L)$ by
\begin{eqnarray}
\tan \psi &=& -\frac{\hat{v}\cdot{\bf p}}{\hat{u}\cdot{\bf p}}
= \frac{\hat{L}\cdot \hat{u} }{\hat{L}\cdot \hat{v}} \nonumber \\
&=& \frac{\cos\theta\cos(\phi-\phi_L)\sin\theta_L -\cos\theta_L\sin\theta}{\sin\theta_L \sin(\phi-\phi_L)} \, ,
\end{eqnarray}
where we have used
\begin{equation}
{\bf p} = \hat{n} \times \hat{L}.
\end{equation}
The parameters $\kappa$ and $\lambda$ define the initial location and orientation of the LISA
constellation. They are related to the quantities
$\bar\phi_0$ and $\alpha_0$ in Cutler's\cite{cc}
equations (3.3) and (3.6) according to $\kappa = \bar\phi_0$ and
$\lambda = 3\pi/4 +\bar\phi_0 - \alpha_0$.
Our compact expression (\ref{lf}) for the low frequency limit agrees with Cutler's\cite{cc}
result, but the agreement is by no means obvious. The equality can be
established using a computer algebra program or by direct numerical evaluation.

\section{Spectral noise}

The variation in the optical path length will be reflected in the noise transfer functions.
For example, the noise $n^s_{21}(t)$ enters into the $X$ variable as
\begin{equation}
N(t)= n^s_{21}(t)-n^s_{21}(t_1) \, .
\end{equation}
Writing $n^s_{21}(t)=n(t)$ and assuming that the armlengths are fixed,
$\ell_{31}(t_1)=\ell_{31}(t_3)=L$, yields the standard result
\begin{equation}
N(f)=n(f)\left(1-e^{2if/f_*}\right),
\end{equation}
and
\begin{equation}
S_N(f) = 4 \sin^2\left(\frac{f}{f_*}\right)S_n(f) .
\end{equation}
Here $S_n(f)$ is the noise spectral density in the photodetector and we have used
the assumption
\begin{equation}
<\! n(f) n^*(f') \! > = \delta(f-f') S_n(f) \, ,
\end{equation}
where the brackets $<>$ denote an ensemble average. The situation is much
more complicated when the armlengths vary since
\begin{equation}
N(f)=n(f)-\frac{1}{2\pi}\int n\bigl(t_1(t)\bigr) e^{2\pi i f t} dt \, .
\end{equation}
Because $\ell(t)$ varies with a one year period, the transfer function will develop sidebands
at $f\pm n f_m$ where $n$ takes integer values. Working to zeroth order in $v$ and lowest
order in $e$ we have
\begin{equation}\label{t1}
t_1(t) \simeq t- 2 L_{12}(t) \, .
\end{equation}
Using the expression (\ref{l12}) for $L_{12}(t)$ and the expansion
\begin{equation}
e^{i x \sin(2\pi f_m t)} = \sum_{k=-\infty}^{\infty} J_k(x) e^{2\pi i f_m k t},
\end{equation}
where $J_k$ is a Bessel function of the first kind of order $k$, allows us to
write
\begin{eqnarray}
&&N(f)=n(f)-\sum_{j=-\infty}^{\infty}\sum_{k=-\infty}^{\infty} n(f+(j+3k)f_m)e^{2if/f_*}
\nonumber \\
&& \; \times e^{2if_m(j+3k)/f_*} e^{i\pi(j-3k)/6}J_{j}\left(\frac{15 e f}{16 f_*}\right)
J_{k}\left(\frac{e f}{16 f_*}\right) .
\end{eqnarray}
The dependence on the Bessel functions tells us that the sidebands only become
significant for frequencies approaching $f_*/e \approx 1$ Hz.
Below the transfer frequency $f_* \sim 10$ mHz it is safe to ignore
the time variation of the arm lengths in calculations
of the noise transfer functions.

\section{Discussion}

We have shown that our general expression for the LISA response function reproduces the
standard static and low frequency limits. However, we have said little about how the
general result should be used. Given a specific model for the orbit, such as the simple
Keplerian model described in the Appendix, it is possible to solve the implicit relations
for the detector orientation, arm lengths and emission times, as we did
in equations (\ref{el12}), (\ref{l12}) and (\ref{t1}). These can then be used to
give explicit expressions for the Michelson or $X$ variables. We did not quote these
expressions as they are very large and not very informative. Ultimately, any application
that requires the full LISA response function is likely to be numerical. It is a simple
matter to write a computer program that returns the LISA response function using equations
(\ref{x1}), (\ref{phs}) and (\ref{main}). If one is only interested in sources with
frequencies below $\sim 5$ mHz, the low frequency approximation (\ref{lf}) will suffice,
but for accurate astrophysical parameter estimation above 5 mHz, the full
LISA response function has to be used.

\section*{Acknowledgements}

We thank Ron Hellings for many help discussions. This work
was supported by the NASA EPSCoR program through Cooperative Agreement NCC5-579.

\section*{Appendix: Keplerian Spacecraft Orbits}

For a constellation of spacecraft in individual Keplerian orbits with an inclination of
$i = \sqrt{3}e$ the coordinates of each spacecraft are given by the expressions
\begin{eqnarray} \label{polcor}
x &=& r\left(\cos(\sqrt{3} e)\cos\beta\cos\gamma -\sin\beta\sin\gamma\right) \nonumber \\
y &=& r\left(\cos(\sqrt{3} e)\sin\beta\cos\gamma +\cos\beta\sin\gamma\right) \nonumber \\
z &=& -r \sin(\sqrt{3} e)\cos\gamma\, .
\end{eqnarray}
where $\beta = 2n\pi/3+\lambda$ ($n=0,1,2$) is the relative orbital phase of each spacecraft in the
constellation, $\gamma$ is the ecliptic azimuthal angle, and $r$ is the standard Keplerian
radius
\begin{equation}
r = \frac{a(1-e^2)}{1+e\cos\gamma} \, .
\end{equation}
Here $a$ is the semi-major axis of the guiding center and has an approximate value of one
AU.

To get the above coordinates as a function of time we first note that the azimuthal angle is
related to the eccentric anomaly, $\psi$, by
\begin{equation} \label{ecan}
\tan\left(\frac{\gamma}{2}\right) = \sqrt{\frac{1+e}{1-e}}
\tan\left(\frac{\psi}{2}\right) \, ,
\end{equation}
and the eccentric anomaly is related to the orbital phase $\alpha(t)=2\pi f_m t + \kappa$ through
\begin{equation}\label{kep}
\alpha - \beta = \psi - e \sin\psi \, .
\end{equation}
For small eccentricities we can expand equations
(\ref{ecan}) and (\ref{kep}) in a power series in $e$ to arrive at
\begin{equation}
\gamma = (\alpha-\beta) + 2 e \sin(\alpha-\beta) + \frac{5}{2}e^2 \sin(\alpha-\beta)
\cos(\alpha-\beta) + \dots
\end{equation}
Substituting this series into equation (\ref{polcor}) and keeping terms only up to order
$e$ gives us
\begin{eqnarray}
x &=& a\cos(\alpha) + ae\left(\sin\alpha\cos\alpha\sin\beta
-(1+\sin^2\alpha)\cos\beta\right)\nonumber \\
y &=& a\sin(\alpha) + ae\left(\sin\alpha\cos\alpha\cos\beta
-(1+\cos^2\alpha)\sin\beta\right)\nonumber \\
z & = & -\sqrt{3} a e \cos(\alpha-\beta) \, .
\end{eqnarray}
These are the desired coordinates of each spacecraft as a function of time. Notice that
by keeping only linear terms in the eccentricity we are neglecting the variation in the
optical path length. The path length will change due to the Keplerian orbits, but these
effects enter at ${\cal O}(e^2)$ and above.


\begin{thebibliography}{99}
\bibitem{cc} C. Cutler, Phys. Rev. D {\bf 57}, 7089 (1998).
\bibitem{aet} M. Tinto, J. W. Armstrong \& F. B. Estabrook,
Phys. Rev. D{\bf 63}, 021101(R) (2001).
\bibitem{lhh} S. L. Larson, W. A. Hiscock, \& R. W. Hellings, Phys.
Rev. D {\bf 62}, 062001 (2000).
\bibitem{cl} N. J. Cornish \& S. L. Larson,  Class. Quant. Grav. {\bf 18} 3473 (2001)
\bibitem{mh} T. A. Moore \& R. W. Hellings, Phys. Rev. D {\bf 65},
062001 (2002).
\bibitem{lppar} P. L. Bender \textit{et al}., LISA Pre-Phase A
Report, 1998.
\bibitem{njc} N. J. Cornish, Class. Quant. Grav. {\bf 18} 4277 (2001).
\bibitem{ron} R. W. Hellings, Phys. Rev. D {\bf 17} 832 (1981).
\bibitem{ta} M. Tinto \& J. W. Armstrong, Phys. Rev. D {\bf 59}, 102003 (1999).
\bibitem{kip} K. S. Thorne, in {\em 300 Years of Gravitation}, edited by S.W. Hawking
and W. Israel (Cambridge University Press, Cambridge, England, 1987), pp. 330-458.
\end{thebibliography}
\end{document}